\documentclass{PoS}
\usepackage{cite}

\title{The critical line of QCD with four degenerate quarks}

\ShortTitle{Critical line of QCD}

\author{P. Cea\\
            Dipartimento di Fisica dell'Universit\`a di Bari, I-70126 Bari, Italy and INFN, Sezione di Bari, I-70126 Bari, Italy\\
            E-mail: \email{paolo.cea@ba.infn.it}}
           
\author{\speaker{L. Cosmai}\\
        INFN, Sezione di Bari, I-70126 Bari, Italy\\
        E-mail: \email{leonardo.cosmai@ba.infn.it}}
            
\author{M. D'Elia\\
            Dipartimento di Fisica dell'Universit\`a di Genova, I-16146 Genova, Italy and INFN, Sezione di Genova, I-16146 Genova, Italy\\
            E-mail: \email{massimo.delia@ge.infn.it}}

\author{A. Papa\\
            Dipartimento di Fisica dell'Universit\`a della Calabria, I-87036 Rende (Cosenza), Italy and INFN, Gruppo collegato di Cosenza, I-87036 Rende (Cosenza), Italy\\
            E-mail: \email{papa@cs.infn.it}}

\abstract{We determine the pseudo-critical couplings at imaginary chemical potentials by high-statistics Monte Carlo simulations of QCD with four degenerate quarks at non-zero temperature and baryon density by the method of analytic continuationan. We reveal deviations from the simple quadratic dependence on the chemical potential visible in earlier works on the same subject. Finally, we discuss the implications of our findings for the shape of the pseudo-critical line at real chemical potential, comparing different possible extrapolations.}

\FullConference{The XXVIII International Symposium on Lattice Field Theory, Lattice2010\\
		June 14-19, 2010\\
		Villasimius, Italy}

\begin{document}

\section{Introduction}
Understanding the phase diagram of QCD on the temperature-chemical potential $(T,\mu)$ has many important implications in cosmology, in astrophysics and in the phenomenology of heavy ion collisions.
Unfortunately, the study of QCD at nonzero baryonic density by numerical simulations on a space-time lattice is plagued by the well-known sign problem: the fermion determinant is complex and the Monte Carlo sampling becomes unfeasible~\cite{Philipsen:2010gj}.

One of the possibilities to circumvent the sign problem is to perform Monte Carlo numerical simulations for imaginary values of the chemical potential, where the fermion determinant is real and the sign problem is absent, and to infer the behavior at real chemical  potential by analytic continuation.
The idea of formulating a theory at imaginary chemical potential $\mu$ was first suggested in Ref.~\cite{Alford:1998sd}. Soon after there were first applications 
to QCD~\cite{Lombardo:1999cz,deForcrand:2002yi,deForcrand:2003hx,D'Elia:2002gd,D'Elia:2004at}.   

The method of analytic continuation has several advantages, indeed coupling $\beta$ and chemical potential $\mu$ can be varied independently and there is 
no limitation for increasing lattice sizes. But  the extent of the attainable domain with real  $\mu$ is limited by the periodicity and non-analyticities~\cite{Roberge:1986mm} and by the accuracy of the interpolation of data for imaginary $\mu$.

Analytic continuation can be exploited for evaluating physical observables at real chemical potential. Indeed a careful numerical analysis in SU(2) has shown that a considerable improvement can be achieved if ratio of polynomials are used as interpolating function~\cite{Cea:2006yd}. 
However the main goal of the application of the method of analytic continuation is locating the critical line on the $(T,\mu)$-plane for real $\mu$.

In previous studies~\cite{Cea:2007vt,Cea:2009ba}  we investigated two-colors QCD at finite baryon chemical potential and  QCD at finite isospin chemical potential. These theories are free from the sign problem and simulations can be performed both at real and imaginary chemical potential. In this way it is possible to check the reliability of the analytic continuation. 
The lesson we learned from the aforementioned studies was that the prediction for the pseudocritical  couplings at real chemical potentials may be wrong if data  at imaginary       $ \mu$  are fitted according to a linear dependence.

We present here results obtained in the determination of the  pseudocritical line  in SU(3)  with $N_f=4$  at finite baryon density~\cite{Cea:2010md}.

\section{Numerical results}

We have considered consider QCD with $N_f=4$  degenerate standard staggered fermions  of mass $am=0.05$. 
Lattice simulations have been performed on a $12^3 \times 4$ lattice 
using the exact $\Phi$  algorithm~\cite{Gottlieb:1987mq},  properly modified  for the inclusion of a finite chemical potential. 
Typical statistics collected is  10k trajectories of 1 molecular dynamics unit, growing 
up to 100k trajectories for the values of the couplings around  the peak of the susceptibility of a given observable.

In order to obtain the critical line on the $(T,\mu)$-plane for real $\mu$ we first need to locate the (pseudo-)critical couplings for several fixed values of the imaginary chemical potential. This has been done by looking for peaks in the susceptibilities of a given observable. A suitable interpolation of the (pseudo-)critical couplings at imaginary chemical potential is then extrapolated to real chemical potential. 

\begin{figure}[htbp]
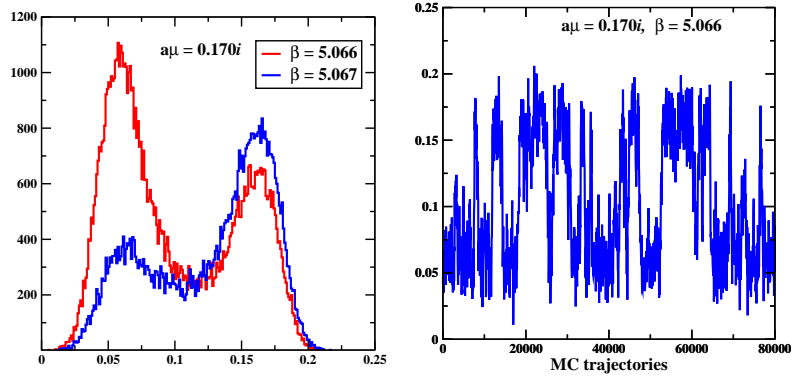
 
\centering 
\begin{minipage}[c]{.35\textwidth}
\centering
\includegraphics[width=.95\textwidth,clip]{histo_muimm_0.170-COLORS.eps} 
\end{minipage}%
\hspace{1mm}%
\begin{minipage}[c]{.35\textwidth}
\centering
\includegraphics[width=.95\textwidth,clip]{history_muimm_0.170_b_5.066-COLORS.eps} 
\end{minipage} 
\caption{(Left) Distribution of the real part of the Polyakov loop 
in SU(3) with $N_f=4$ on a 12$^3\times 4$ lattice with $am$=0.05 
at $a\mu=0.170 i$ and for two $\beta$ values around the transition.
(Right) Monte Carlo history of the real part of the Polyakov loop 
in SU(3) with $N_f=4$ on a 12$^3\times 4$ lattice with 
$am$=0.05 at $a\mu=0.170i$ and $\beta$=5.066.
\label{fig1}} 
\end{figure}

In SU(3)  with $N_f=4$    the critical line is a line of first  order transitions in the first Roberge-Weiss sector  $-(\pi/3)^2 \leq (\mu/T)^2 \leq 0$. Therefore tunneling between 
the different phases is expected every few thousands trajectories (Fig.~\ref{fig1}).

\begin{figure}[htbp] 
\centering
\includegraphics[width=.35\textwidth,clip]{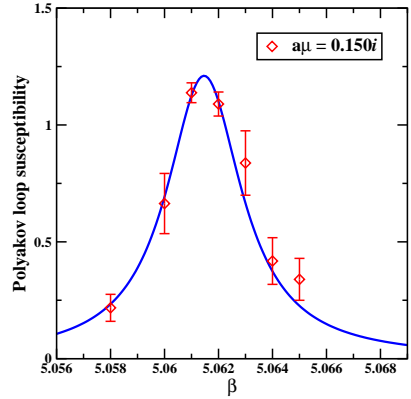}
\caption{Susceptibility of the (real part of the) Polyakov loop
{\it vs} $\beta$ in SU(3) with $N_f=4$ on a 12$^3\times 4$ lattice with 
$am$=0.05 and $a\mu = 0.150i$. The solid lines represent the Lorentzian 
interpolation.}
\label{fig2}
\end{figure}

\begin{figure}[htbp] 
\centering
\includegraphics[width=.35\textwidth,clip]{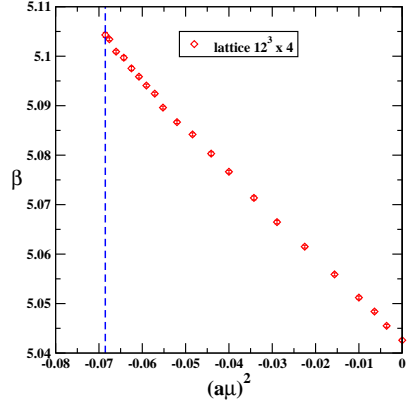}
\caption{Critical couplings obtained in SU(3) with $N_f=4$ 
on a 12$^3\times 4$ lattice with $am$=0.05. The dashed vertical line
indicates the boundary of the first RW sector, $a\, {\rm Im}(\mu)=\pi/12$.}
\label{fig3}
\end{figure}

The critical coupling $\beta(\mu^2)$ at imaginary chemical potential $\mu$  is determined as the value for which the susceptibility of (the real part of) the Polyakov loop exhibits a peak (Fig.~\ref{fig2}). The peak value at given imaginary chemical potential has been determined through a Lorentzian fit. 
We also checked the consistency of our determinations by means of Ferrenberg-Swendsen method or by estimating the point where the peaks in the distribution of 
the real part of the Polyakov loop  has equal height. Moreover we  verified the determination of the pseudo-critical couplings do not change if other  observables are used to perform
the aforementioned analyses.

In Fig.~\ref{fig3} the values of the critical couplings $\beta(\mu^2)$ versus $a \mu^2$ are displayed. Data do not line up along a straight line and 
$\beta(\mu^2)$  cannot be parametrized by a polynomial of order $\mu^2$. Therefore we  are lead to use the following fit function:
\begin{equation}
\label{polyratio}
\frac{a_0+a_1 (a \mu)^2 + a_2 (a \mu)^4 + a_3 (a \mu)^6}{1+a_4 (a
  \mu)^2 + a_5 (a \mu)^4}  \,.
\end{equation}
We found that a $\chi^2/\rm{dof}  \sim 1$ is obtained both with sextic polynomial in $\mu$ (Fig.~4 (left)) and
ratio of polynomials  2nd to 4th order or 4th to 2nd order (Fig.~4 (center)).

\begin{figure}[htbp]
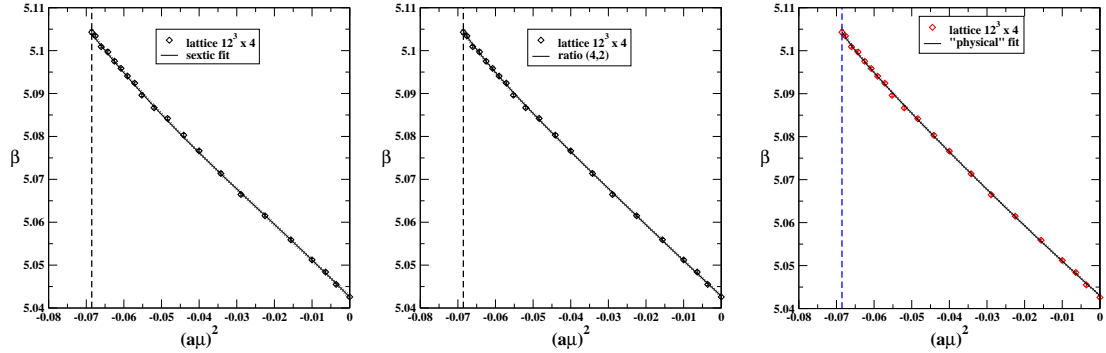

\centering 
\begin{minipage}[c]{.32\textwidth}
\centering
\includegraphics[width=.95\textwidth,clip]{fit_6.eps} 
\end{minipage}%
\hspace{1mm}%
\begin{minipage}[c]{.32\textwidth}
\centering
\includegraphics[width=.95\textwidth,clip]{fit_ratio_4-2.eps} 
\end{minipage} 
\hspace{1mm}%
\begin{minipage}[c]{.32\textwidth}
\centering
\includegraphics[width=.95\textwidth,clip]{fit_physical_2-COLOR.eps} 
\end{minipage} 
\caption{Fits to the critical couplings: (left) plain 6th order polynomial, (center) ratio of a 4th to 2nd order 
polynomial  and (right) physical fit according to Eq.~(\protect\ref{a2beta}).}
\label{fig4}
\end{figure}
\begin{figure}[htbp] 
\centering
\includegraphics[width=.5\textwidth,clip]{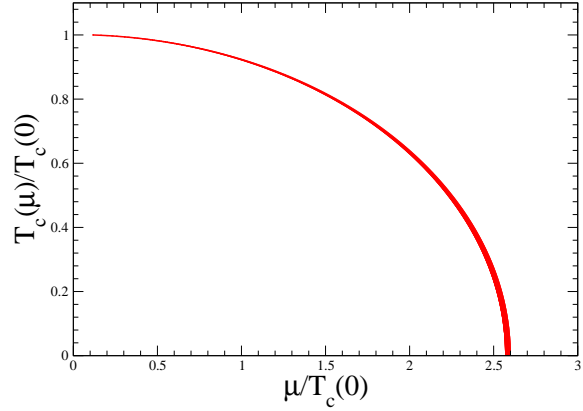}
\caption{The phase diagram obtained extrapolating the critical line down to $T=0$  axis by means
of Eq.~(\protect\ref{phys}).}
\label{fig7}
\end{figure}
\begin{figure}[htbp] 
\centering
\includegraphics[width=.5\textwidth,clip]{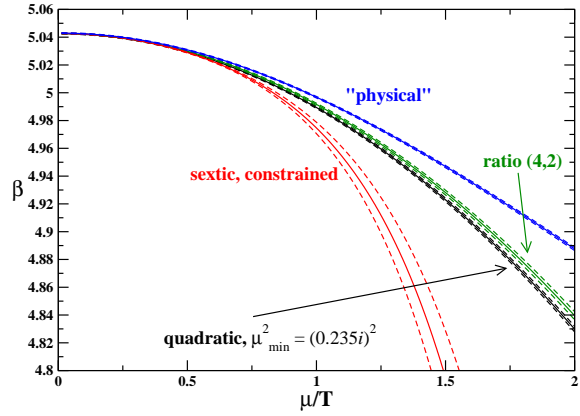}
\caption{Extrapolation to real chemical potentials of the quadratic 
(with $(a\mu)^2 = (0.235i)^2$), sextic constrained, ratio of polynomials 
and ``physical'' fits.}
\label{fig7b}
\end{figure}
\begin{figure}[htbp] 
\centering
\includegraphics[width=.52\textwidth,clip]{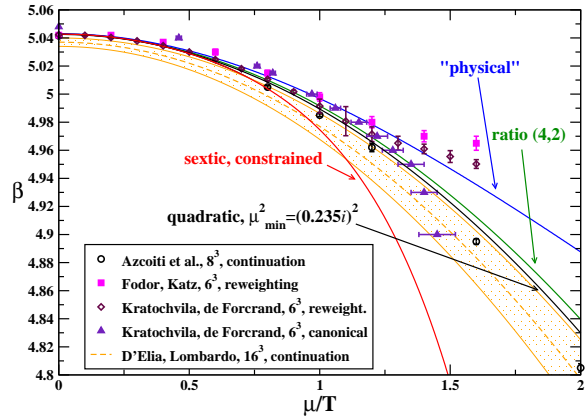}
\caption{Comparison of our extrapolations with other determinations in the
literature. For the sake of readability, our extrapolations have been plotted
without error bands and labels, since they can be easily recovered
from the previous figure. {\it Legenda}: 
D'Elia, Lombardo, Ref.~\cite{D'Elia:2002gd,D'Elia:2004at}; 
Azcoiti {\it et al.}, Ref.~\cite{Azcoiti:2005tv}; 
Fodor, Katz, Ref.~\cite{Fodor:2001au}; 
Kratochvila, de Forcrand, Ref.~\cite{Kratochvila:2005mk}.}
\label{fig7c}
\end{figure}
In addition we  implementd a new fit strategy which consists in 
writing down the interpolating function in physical units ("physical" fit in Fig.~4 (right))
\begin{eqnarray}
\label{phys}
\left[\frac{T_c(\mu)}{T_c(0)}\right]^2=\frac{1+C\mu^2/T_c^2(\mu)}
{1+A\mu^2/T_c^2(\mu)+B\mu^4/T_c^4(\mu)}  \, .
\end{eqnarray}
with $T_c=1/(a(\beta) L_t)$ and $A$, $B$, $C$ fit parameters. 
The implicit relation between $\beta_c$ and $\mu^2$  is given by
\begin{eqnarray}
\label{a2beta}
a^2(\beta_c(\mu^2))|_{\rm 2-loop} = a^2(\beta_c(0))|_{\rm 2-loop}
\times& \frac{1+A\mu^2/T_c^2 + B\mu^4/T_c^4}{1+C\mu^2/T_c^2}  \,.
\end{eqnarray}
Also for the "physical" fit we get a $\chi^2/\rm{dof}  \sim 1$.
Moreover we have extrapolated the critical line  down to $T=0$  axis, under the assumption 
that the physical fit gives the correct behavior of the critical line at real $\mu$ down to $T=0$ (Fig.~(\ref{fig7})).
We obtain the following estimate of the critical value of $\mu$ on the $T=0$ axis:
\begin{equation}
\mu = \sqrt{C/B} \,\, T_c(0)  = 2.5904(93)  \,\,  T_c(0) \,.
\end{equation}

We have so far discussed several interpolations of the data in the $\mu^2 \le 0$ region. A crucial question
is to ascertain if the related extrapolations  to $\mu^2 > 0$ are consistent between them.
Unfortunately, as one can inspect in Fig.~6, different interpolations lead to somewhat distinct extrapolations 
and, unless an extra-argument is found to make one fitting function preferable with respect to the others, 
one cannot rely on a unique extrapolation, except  in the region $\mu/T \le 0.6$. In Fig.~7 several determinations of the critical line existing in the literature 
are presented together our results. 
Looking at Fig.~7, one could 
comment that the extrapolation of the ``physical'' fit exhibits the same trend 
as data from reweighting, whereas that from the sextic constrained fit mimics 
the strong coupling behavior~\cite{Miura:2009nu}, the other two extrapolations 
of ours lying in-between. However, previous determinations at real $\mu$ 
in the literature seem to be in fair agreement up to $\mu/T \simeq 1.2$.
If one takes this common trend as benchmark for our extrapolations,
the ``physical'' and the polynomial ratio (4,2) seem to be favoured.

We have tried to include in our fit also  data at real chemical
potential available from the literature (see Fig.~7).
A serious limitation of this combined approach is the inhomogeneity
of the data presently available, due to different lattices and systematics.
However, if the inhomogeneity of data at real
$\mu$ will be reduced by new Monte Carlo determinations, the combined-fit 
strategy could bring along an appreciable improvement.

\section{Conclusions}
We have revisited the application of the method of analytic continuation from imaginary to real chemical potential in QCD with $N_f=4$ degenerate flavors. The aims were:
\begin{itemize}
\item  to determine precisely the pseudo-critical line $\beta_c(\mu^2)$  in  the region of negative $\mu^2$  (20 data points almost  uniformly distributed in the region $-(\pi/12)^2 \le (a\mu)^2  \le 0)$;
\item  to exploit  interpolating functions sensitive to possible deviations of  the critical line from the quadratic behavior in  $\mu$  for larger absolute  values of $\mu$  (these deviations were clearly  seen in QCD-like theories, such as 2-color QCD and finite isospin QCD, where  it was given compelling evidence that their neglect could mislead the  analytic continuation to real chemical potential);
\item to extrapolate the newly adopted interpolations to the region of real  $\mu$ and to re-determine, therefore, the critical line in QCD.
\end{itemize}

We found that deviations from the quadratic behavior in $\mu$  of  $\beta_c(\mu^2)$  at negative $\mu^2$  are visible in QCD with $N_f=4$.
Several kinds of functions able to interpolate them lead to extrapolations to real $\mu$ which start diverging from each other for $\mu/T \ge 0.6$.
The shortcomings of the method of analytic continuation could be less severe for $N_f=2$ or $N_f=2+1$ (where sensitivity to nonlinear terms in $\mu^2$ could be enhanced).
Moreover possible improvement could come by theoretical development able to discriminate between interpolations, or by a combined numerical strategy aimed at gathering information from different approaches (such as reweighting, canonical approach, etc.)  applied so far independently from each other.


\providecommand{\href}[2]{#2}\begingroup\raggedright\endgroup



\end{document}